\documentclass[showkeys]{revtex4}

\usepackage{graphicx}
\usepackage{amsmath,amsthm,amssymb,bm,amsfonts}
\usepackage{todonotes}

\def\half{{\textstyle{\frac{1}{2}}}}

\def\qbar{{\overline{q}}}

\def\qbold{{\mathbf{q}}}

\def\Nbar{{\bar{N}}}

\def\alfabar{{\bar{\alpha}}}
\def\betabar{{\bar{\beta}}}

\def\Dtilde{\tilde{D}}

\newcommand{\bea}{\begin{eqnarray}}
\newcommand{\eea}{\end{eqnarray}}
\newcommand{\be}{\begin{equation}}
\newcommand{\ee}{\end{equation}}

\preprint{INT-PUB-15-035}
\begin{document}
{INT-PUB-15-035}\vspace*{1.5cm}
\title{\bf Constraining the double gluon distribution by the  single gluon distribution}

\author{Krzysztof Golec-Biernat}
\affiliation{Institute of Nuclear Physics Polish Academy of Sciences, 31-342 Cracow, Poland}
\affiliation{Faculty of Mathematics and Natural Sciences, University of Rzesz\'ow,  35-959 Rzesz\'ow, Poland}

\author{Emilia Lewandowska}
\affiliation{Institute of Nuclear Physics Polish Academy of Sciences, 31-342 Cracow, Poland}

\author{Mirko Serino}
\affiliation{Institute of Nuclear Physics Polish Academy of Sciences, 31-342 Cracow, Poland}

\author{Zachary Snyder}
\affiliation{Penn State University, University Park, PA 16802, United States}

\author{Anna M. Sta\'sto}
\affiliation{Penn State University, University Park, PA 16802, United States}
\affiliation{Institute of Nuclear Physics Polish Academy of Sciences, 31-342 Cracow, Poland}

\begin{abstract}
We show how to consistently construct initial conditions for the QCD evolution equations 
for double parton distribution functions in the pure gluon case. We use to momentum sum  rule for this purpose and a specific
form of the known single gluon distribution function in the MSTW parameterization. The resulting double gluon distribution satisfies exactly the momentum sum rule and is parameter free.
We also study numerically its evolution with a hard scale and show the approximate factorization into product of two single gluon distributions at small values of $x$, whereas at large values of $x$ the factorization is always violated in agreement with the sum rule.
\end{abstract}


\maketitle
\section{Introduction}

Multiparton interactions play an important role in the hadronic collisions at high energies. They occur when at one encounter of the initial hadrons, more than one partonic interaction occurs.  They were first observed and measured at the Tevatron \cite{Akesson:1986iv,Abe:1997bp,Abe:1997xk,Abazov:2009gc} and subsequently a systematic experimental study was performed at the higher energy Large Hadron Collider \cite{Aad:2013bjm,Chatrchyan:2013xxa,Aad:2014rua}.
 The theoretical description of such interactions within perturbative QCD is possible in the presence of the sufficiently hard scales.  The computation 
 of double parton scattering (DPS)
cross sections within the collinear framework makes use of the double parton distribution functions (DPDFs) \cite{Shelest:1982dg,Zinovev:1982be,Ellis:1982cd,Bukhvostov:1985rn,
Snigirev:2003cq,Korotkikh:2004bz,Gaunt:2009re,Blok:2010ge,Ceccopieri:2010kg,Diehl:2011tt,
Gaunt:2011xd,Ryskin:2011kk,Bartels:2011qi,Blok:2011bu,Diehl:2011yj,Luszczak:2011zp,
Manohar:2012jr,Ryskin:2012qx,Gaunt:2012dd,Blok:2013bpa,
vanHameren:2014ava, Maciula:2014pla, Snigirev:2014eua, 
Golec-Biernat:2014nsa, 
 Gaunt:2014rua, Harland-Lang:2014efa,
  Blok:2014rza, Maciula:2015vza}.
In the collinear leading logarithmic approximation DPDFs obey QCD evolution equations 
\cite{Kirschner:1979im,Shelest:1982dg,Zinovev:1982be,Snigirev:2003cq,Korotkikh:2004bz,Gaunt:2011xd}, 
similar to the Dokshitzer-Gribov-Lipatov-Altarelli-Parisi (DGLAP)  evolution equations for single parton distribution functions (PDFs). The evolution equations for DPDFs conserve new sum rules which relate
the double and single parton distributions once they are imposed on initial conditions for the evolution equations at some initial scale. 
All the attempts up till now  to  construct  conditions which satisfy these sum rules were rather unsuccessful, see e.g.
Refs.~\cite{Korotkikh:2004bz,Gaunt:2011xd,Golec-Biernat:2014bva} with an exception of the analysis  \cite{Broniowski:2013xba} for valence quarks only.

In this letter,  we show how to  consistently perform such a  construction   in a pure gluon case, using the known single PDFs in the MSTW parameterization   \cite{Martin:2009iq} 
and the momentum sum rule. We find the parameter free double gluon distribution   which we evolve with our numerical program. In particular, we study 
the build up of its approximately factorizable form for small values of parton momentum fractions, $x_{1,2}<0.1$. The full case with quarks and gluons is postponed to a separate publication.

\section{Evolution equations and sum rules}

We consider the DPDFs with equal hard scales, $Q_1=Q_2\equiv Q$, and the relative momentum $\qbold=0$: 
\be
D_{f_1f_2}(x_1,x_2,Q)\,\equiv\,D_{f_1f_2}(x_1,x_2,Q,Q,\qbold=0)\; ,
\label{eq:dpdfdef}
\ee
where $x_{1,2}\in [0,1]$ are parton momentum fractions, which obey the condition $x_1+x_2\le1$,  and $f_{1,2}$ are parton flavors (including gluon)
\cite{Diehl:2011tt,Diehl:2011yj}. In this case, the 
evolution equations in the  leading logarithmic approximation read 
\begin{align}\nonumber
\label{eq:twopdfeq}
\frac{\partial}{\partial {\ln Q^2}}\, D_{f_1f_2}(x_1,x_2,Q)
=\frac{\alpha_s(Q)}{2\pi}\sum_{f'}
&\Bigg\{\int^{1-x_2}_{x_1} \frac{du}{u} \,{\cal{P}}_{f_1f'}\!\left(\frac{x_1}{u}\right) D_{f' f_2}(u,x_2,Q)
\\\nonumber
&+\int_{x_2}^{1-x_1} \frac{du}{u}\,{\cal{P}}_{f_2f'}\!\left(\frac{x_2}{u}\right)D_{f_1f'}(x_1,u,Q)
\\
&+ \frac{1}{x_1+x_2}\,{P}^R_{f'\to f_1f_2}\!\left(\frac{x_1}{x_1+x_2}\right) D_{f'}(x_1+x_2,Q)\Bigg\}.
\end{align}
where
the functions  $\cal{P}$ on the r.h.s. are  the leading order Altarelli-Parisi splitting functions (with virtual corrections for ${\cal{P}}_{ff}$ included).  The third term on the r.h.s corresponds to the splitting of one parton into two daughter partons, described by the Altarelli-Parisi splitting function for real emission, $P^R_{f'\to f_1f_2}$. It contains the single PDFs, $D_{f'}$, thus eq.~\eqref{eq:twopdfeq} has to be solved together with the ordinary DGLAP equations, see e.g. Ref.~\cite{Gaunt:2011xd} for more details.

The significance of the splitting terms in the evolution equations \eqref{eq:twopdfeq} for the computation of the double parton scattering cross sections was a subject of intensive debate in the literature over the last few years  \cite{Diehl:2011tt,Gaunt:2011xd, 
Blok:2011bu,Diehl:2011yj,Manohar:2012pe,Gaunt:2012dd,Golec-Biernat:2014nsa,Gaunt:2014rua}.  The conclusion which emerges from this discussion is that the processes which are summed up by the splitting terms and coming from both hadrons
in hadron-hadron collisions should rather be classified as the single parton scattering process \cite{Gaunt:2012dd}. On the other hand, the so called single splitting contributions,  with parton splitting  from one hadronic side only, are important for
the double parton scattering cross sections \cite{Blok:2010ge,Blok:2012jr,Golec-Biernat:2014nsa,Gaunt:2014rua}. 
From the perspective of the present paper, in which we only concentrate on the evolution of the DPDFs, the splitting
terms in the evolution equations are crucial for the conservation of sum rules which are discussed below.

The sum rules which are conserved by the  evolution equations (\ref{eq:twopdfeq})  are the momentum and valence quark
number sum rules \cite{Gaunt:2009re}. Imposing them for initial conditions specified at some initial scale $Q_0$, 
they are guaranteed to be satisfied at any other scale $Q$.  
The momentum sum rule for the DPDFs reads
\begin{eqnarray}
\label{eq:momrule1}
\sum_{f_1}\int_{0}^{1-x_2}dx_1\,x_1D_{f_1f_2}(x_1,x_2)=(1-x_2)D_{f_2}(x_2) \; ,
\end{eqnarray}
while the valence  quark number sum rule is given by
\begin{eqnarray}
\label{eq:valrule1}
\int_0^{1-x_2}dx_1\!\left\{D_{qf_2}(x_1,x_2)-D_{\qbar f_2}(x_1,x_2)\right\} =
(N_q-\delta_{f_2q}+\delta_{f_2\qbar}) D_{f_2}(x_2)\,,
\end{eqnarray}
where $q=u,d,s$ and  $N_u=2, N_d=1,N_s=0$ are the valence quark number for each of the quark flavors. 
The same relations hold true with respect to the second parton 
\begin{align}
\label{eq:momrule2}
\sum_{f_2}\int_{0}^{1-x_1}dx_2\,x_2D_{f_1f_2}(x_1,x_2) &=(1-x_1)D_{f_1}(x_1)\; ,
\\
\int_0^{1-x_1}dx_2\!\left\{D_{f_1 q}(x_1,x_2)-D_{f_1\qbar}(x_1,x_2)\right\} &=
(N_q-\delta_{f_1q}+\delta_{f_1\qbar}) D_{f_1}(x_1)\,.
\label{eq:valrule2}
\end{align}
Notice that if the DPDFs are  parton exchange symmetric, 
\be
\label{eq:pes}
D_{f_1f_2}(x_1,x_2)=D_{f_2f_1}(x_2,x_1)\,,
\ee
the sum rules with respect to the first parton imply the sum rules with respect to the second one since   the evolution equations
also conserve parton exchange symmetry. 

We see that the above sum rules relate the double and single parton distribution
functions, which reflects the common origin of those distributions, namely, the  expansion of the nucleon state in Fock
light-cone components \cite{Gaunt:2009re}. In addition,  the sum rules for the single parton distributions are also satisfied -
the momentum sum rule
\begin{align}
\sum_f\int_0^1dx\,xD_f(x)=1
\end{align}
and the quark valence sum rule for  $q=u,d,s$
\begin{align}
\int_0^1dx\,\left\{D_q(x)-D_{\qbar}(x)\right\}=N_q\,.
\end{align}

\section{Mellin moment formulation}

Let us perform the double Mellin transform  of the  DPDFs
\be
\Dtilde_{f_1f_2}(n_1,n_2) = \int_0^1dx_1\int_0^1dx_2\,  (x_1)^{n_1-1}(x_2)^{n_2-1}
{D}_{f_1f_2}(x_1,x_2) \Theta(1-x_1-x_2).
\label{eq:doublemellin}
\ee
where $n_{1,2}$  are complex numbers and we omit the scale $Q_0$ in the notation from now on.
The step function $\Theta(1-x_1-x_2)$ is inserted into the definition of the Mellin transform since this is the region over which the double parton distribution is defined.  Similarly, for the single parton distribution functions, we define the Mellin moments
\be
\Dtilde_f(n) = \int_0^1dx\,  x^{n-1}D_f(x)\, ,
\label{eq:singlemellin}
\ee
where $n$ is a complex number.  
The Mellin moments  can be transformed  back to the $x$-space using the inverse transformation
for the single parton distribution,
\be
\label{eq:singlemellininv}
D_{f}(x_1) =\int_{C}\frac{dn}{2\pi i}\,(x_1)^{-n}
\,\Dtilde_{f}(n)\, ,
\ee
and similarly for the double parton distribution function
\be
\label{eq:doublemellininv}
D_{f_1f_2}(x_1,x_2) =\int_{C_1}\frac{dn_1}{2\pi i}\,(x_1)^{-n_1}
\int_{C_2}\frac{dn_2}{2\pi i}\,(x_2)^{-n_2}\,\Dtilde_{f_1f_2}(n_1,n_2),
\ee
 where the integration contours $C_1$ and $C_2$ lie to the right of the rightmost singularity in the complex plane of $n_1$ and $n_2$, respectively. Let us emphasize that formula (\ref{eq:doublemellininv}) is only applicable to $x_{1,2}\in[0,1]$
and $x_1+x_2\le 1$. 

The sum rules (\ref{eq:momrule1}) and (\ref{eq:valrule1}) can be written with the help of the Mellin moments  after the  integration of both sides over $x_2$ with the factor $(x_2)^{n_2-1}$. Thus,  we find
\begin{eqnarray}
\label{eq:mom1}
\sum_{f_1}\,\Dtilde_{f_1f_2}(2,n_2) &=& \Dtilde_{f_2}(n_2) - \Dtilde_{f_2}(n_2+1)\, ,
\\\nonumber
\\
\Dtilde_{q f_2}(1,n_2) - \Dtilde_{\bar{q}f_2}(1,n_2) &=& (N_q-\delta_{f_2 q}+\delta_{f_2\qbar}) 
\Dtilde_{f_2}(n_2). 
\label{eq:val1}
\end{eqnarray}
Analogous relations hold  true for the  second parton
\begin{eqnarray}
\label{eq:mom2}
\sum_{f_2}\,\Dtilde_{f_1f_2}(n_1,2) &=& \Dtilde_{f_1}(n_1) - \Dtilde_{f_1}(n_1+1)\,,
\\\nonumber
\\
\Dtilde_{f_1q}(n_1,1) - \Dtilde_{f_1\qbar}(n_1,1) &=& (N_q-\delta_{f_1 q}+\delta_{f_1\qbar}) 
\Dtilde_{f_1}(n_1).
\label{eq:val2}
\end{eqnarray}
 These sum rules have to be satisfied simultaneously with the momentum sum rule for the single parton distribution
 \be
 \label{eq:momsingle}
 \sum_f \tilde{D}_f(2) = 1\; ,
 \ee
 and the valence quark sum rule
 \be
 \label{eq:valsingle}
 \tilde{D}_q(1)-\tilde{D}_\qbar(1)=N_q\,.
 \ee

It would be extremely useful to construct initial conditions for DPDFs which fulfill the above sum rules since the PDFs on the r.h.s of 
Eqs.~(\ref{eq:momrule1})-(\ref{eq:valrule2}) are very well known from the global analysis fits. Thus, the PDFs
constrain the DPDFs, solving or significantly reducing
the problem of uncertainty in the specification of initial conditions for DPDFs evolution.
For this purpose, we consider the single PDF parametrization from the MSTW fits \cite{Martin:2009iq}. We will choose the leading order (LO) version of this
parameterization since the evolution equations (\ref{eq:twopdfeq}) are given in  leading logarithmic approximation. We start from considering the case with only gluons. This limits the set of the possible distributions in  which $D_{gg}$ and $D_g$ are the only relevant functions and of course we only have to fulfill the momentum sum rule.

\section{Pure gluon case}

The single gluon distribution is specified in the LO MSTW parameterization at the scale $Q_0=1~{\rm GeV}$ and is given in the form 
\be\label{eq:mstwg}
 D_g(x) = A_g\,x^{\delta_g-1} (1-x)^{\eta_g} (1 + \epsilon_g\,\sqrt{x} + \gamma_g\,x) \, ,
 \ee
where $A_g=0.0012216, \delta_g=-0.83657, \eta_g=2.3882, \epsilon_g=-38.997$ and $\gamma_g=1445.5$. For errors on these parameters
and the discussion of their determination, see \cite{Martin:2009iq}. 
Since we only use gluons in our analysis, we renormalize the gluon distribution such that the total longitudinal momentum carried  
by gluons equals one, which results in $A_g=0.0033723$. 
This is really not so crucial here as the normalization can be set arbitrarily for the single channel case and it does not affect the subsequent discussion.
The parametrization  (\ref{eq:mstwg}) can be written in a general  form which is more suitable for our purpose
\be\label{eq:newmstwg}
 D_g(x) \; = \; \sum_{k=1}^L N^k_g\, x^{\alpha^k_g} \, (1-x)^{\beta_g^k}\,,
 \ee
 where $L=3$ and the parameters $N^k_g, \alpha^k_g$ and $\beta_g^k$ can easily be  found by the  comparison with eq.~(\ref{eq:mstwg}),~
 \begin{align}\nonumber
 N_g^1 &= A_g\,,~~~~~~~~~~~~~N_g^2 = \epsilon_gA_g\,,~~~~~~~~~~~N_g^3=\gamma_gA_g
 \\\label{eq:parameters}
 \alpha_g^1 &= \delta_g-1\,,~~~~~~~~~\alpha_g^2 = \delta_g-\half\,,~~~~~~~~~\alpha_g^3=\delta_g\,,~~~~~~~~~ \beta_g^1 =\beta_g^2=\beta_g^3=\eta_g\,.
   \end{align}
 In the Mellin space, the gluon distribution \eqref{eq:mstwg} can be written as
\begin{equation}
\tilde{D}_g(n) \; = \;  \sum_{k=1}^{L} N^k_g \, \frac{\Gamma(n+\alpha^k_g)\Gamma(\beta^k_g+1)}{\Gamma(n+\alpha^k_g+\beta^k_g+1)} \; .
\label{eq:dnsumk}
\end{equation}
where the expresion on the r.h.s., $\Gamma(x)\Gamma(y)/\Gamma(x+y)\equiv B(x,y)$, is the  Euler Beta function.  Thus the MSTW parametrization for the initial condition is in the form of the sum over the Beta functions with different sets of 
 parameters which govern the small $x \rightarrow 0$ and large $x\rightarrow 1$ behavior. 

For the double parton distribution 
$D_{gg}$ we shall take the following ansatz in the form 
\be\label{eq:gg1}
D_{gg}(x_1,x_2) \; = \; \sum_{k=1}^L \Nbar_{gg}^k\,(x_1x_2)^{\alfabar_g^k}\,(1-x_1-x_2)^{\betabar_g^k} \, ,
\ee
where $ \Nbar_{gg}^k, \alfabar_g^k$ and $\betabar_g^k$ are the parameters to be determined.
The above ansatz is in the form of the sum over the Dirichlet-type distributions of order $K=3$
\be
\label{eq:dirch}
f(x_1,x_2;\gamma_1,\gamma_2,\gamma_3) = N \prod_{i=1}^{K=3} x_i^{\gamma_i}\, ,
\ee
with
$x_1,x_2 >0, x_1+x_2 \le 1$ and $x_3 = 1-x_1-x_2$.
Notice that the function \eqref{eq:gg1} is parton exchange symmetric, 
$D_{gg}(x_1,x_2)=D_{gg}(x_2,x_1)$,
as it should be.   It should also  fulfill the momentum sum rules with respect to both partons.  
Also note that the number of terms in this sum, $L$,  is the same as the number of terms in the single parton distribution \eqref{eq:newmstwg}.
The Mellin space representation of the above ansatz reads
\begin{equation}
\tilde{D}_{gg}(n_1,n_2) \; = \;  \sum_{k=1}^L  \Nbar_{gg}^k  \frac{\Gamma(n_1+\alfabar_g^k)\Gamma(n_2+\alfabar_g^k)\Gamma(1+\betabar_g^k)}{\Gamma(n_1+n_2+1+\betabar_g^k+2\alfabar_g^k)} \; ,
\label{eq:d2nsumk}
\end{equation}
which is in the form of the generalized Beta function. 

In the pure
 gluon case only the momentum sum rule for the DPDFs in the momentum space  reads, 
\begin{eqnarray}
\label{eq:momrule1g}
\int_{0}^{1-x_2}dx_1x_1D_{gg}(x_1,x_2)=(1-x_2)D_{g}(x_2)\, ,
\end{eqnarray}
and similarly for the momentum sum rule with respect to the second gluon.
In the Mellin representation this condition reduces to
\be
\Dtilde_{gg}(2,n_2) = \Dtilde_{g}(n_2) - \Dtilde_{g}(n_2+1)\, ,
\label{eq:mom2gluon}
\ee
It is easy to see that the distributions of the form presented in Eqs.~\eqref{eq:newmstwg}
and \eqref{eq:gg1} fulfill the momentum sum rule provided certain constraints are satisfied.
The right hand side of Eq.~\eqref{eq:mom2gluon} is  the difference of the moments of  single parton distributions  which can be written as
\begin{equation}
\tilde{D}_g(n_2)-\tilde{D}_g(n_2+1) = \sum_{k=1} ^L N^k_g \, B(n_2+\alpha^k_g,\beta^k_g+2) = \sum_{k=1} ^L N^k_g \frac{\Gamma(n_2+\alpha^k_g)\Gamma(\beta^k_g+2)}{\Gamma(n_2+\alpha^k_g+\beta^k_g+2)} \; ,
\label{eq:rhs}
\end{equation}
where we  used the following property of the Beta function 
\be
B(a,b)=B(a+1,b)+B(a,b+1) \; .
\label{eq:betafun}
\ee
On the other hand the left-hand-side of Eq.~\eqref{eq:mom2gluon} is obtained   upon setting $n_1=2$ in Eq.~\eqref{eq:d2nsumk}
\begin{equation}
\tilde{D}_{gg}(2,n_2) =  \sum_{k=1}^L  \Nbar_{gg}^k  \frac{\Gamma(2+\alfabar_g^k)\Gamma(n_2+\alfabar_g^k)\Gamma(1+\betabar_g^k)}{\Gamma(3+n_2+\betabar_g^k+2\alfabar_g^k)} \; .
\label{eq:lhs}
\end{equation}
Now in order for the momentum sum rule to be satisfied, we need Eqs.~\eqref{eq:rhs} and \eqref{eq:lhs} to be equal term by term in the sum over $k$. From the requirement that the poles and zeros in $n$ in each term should be in the same location we  find that
\be
\alfabar_g^k =\alpha^k_g\,,~~~~~~~~~~ 2\alfabar^k_g+\betabar^k_g+3=\alpha_g^k+\beta_g^k+2\, ,
\ee
and from the requirement that the normalization of each terms should be the same we have that
\be
\Nbar_{gg}^k\,
\Gamma(2+\alfabar_g^k)\Gamma(1+\betabar_g^k)= N^k_g\, \Gamma(2+\beta_g^k)\,.
\ee
From these relations we compute all the parameters of the double gluon distribution in terms of the known parameters of the single gluon distribution, given by
Eq.~\eqref{eq:parameters}, 
 to find the following parameter-free double distribution at the initial scale $Q_0=1~{\rm GeV}$,
\be\label{eq:initialgg}
D_{gg}(x_1,x_2)=\sum_{k=1}^3 \, N^k_g\frac{\Gamma(\beta^k_g+2)}{\Gamma(\alpha^k_g+2)\Gamma(\beta^k_g-\alpha^k_g)}\,
(x_1x_2)^{\alpha^k_g}\,(1-x_1-x_2)^{\beta^k_g-\alpha^k_g-1}\,,
\ee
satisfing the momentum sum rule \eqref{eq:momrule1g} by construction. 
Notice that even for small momentum fractions, $x_{1,2}\ll 1$, the resulting   double gluon distribution is not factorizable, i.e. $D_{gg}(x_1,x_2)\ne D_g(x_1)D_g(x_2)$.

\begin{figure*}[t]
\includegraphics[width = 10cm]{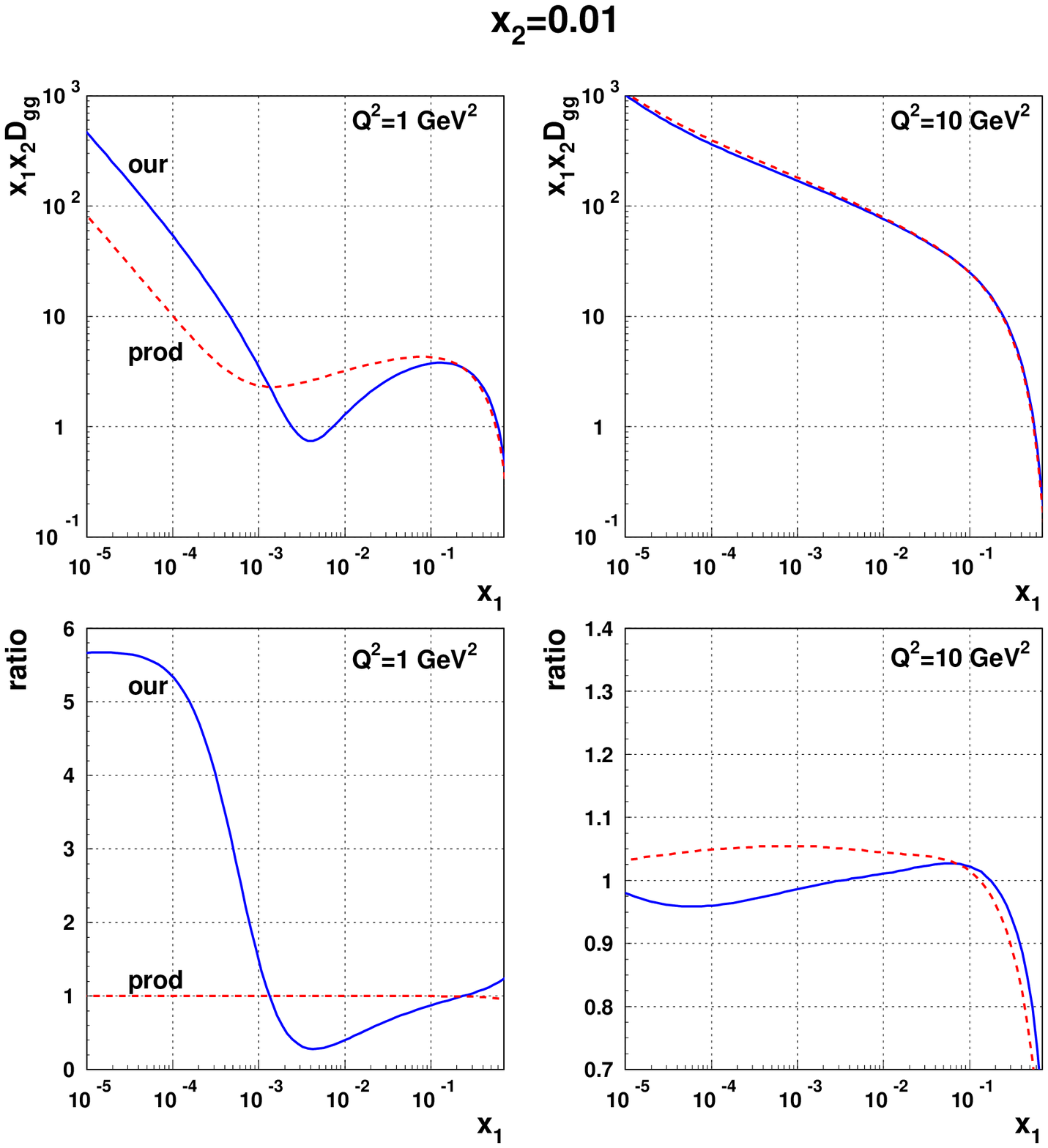}
\caption{The distribution $x_1x_2D_{gg}(x_1,x_2=10^{-2})$ at  $Q_0^2=1~{\rm GeV}^2$ (left upper panel) and 
$Q^2=10~{\rm GeV}^2$ (right upper panel) and the ratio \eqref{eq:ratio} (lower panels). The solid lines correspond to input (\ref{eq:initialgg}) (our)
while the dashed lines to input (\ref{eq:gaunt}) (prod).
}
\label{fig1}
\end{figure*}

\section{Evolution of double gluon distribution}

\begin{figure*}[t]
\includegraphics[width = 10cm]{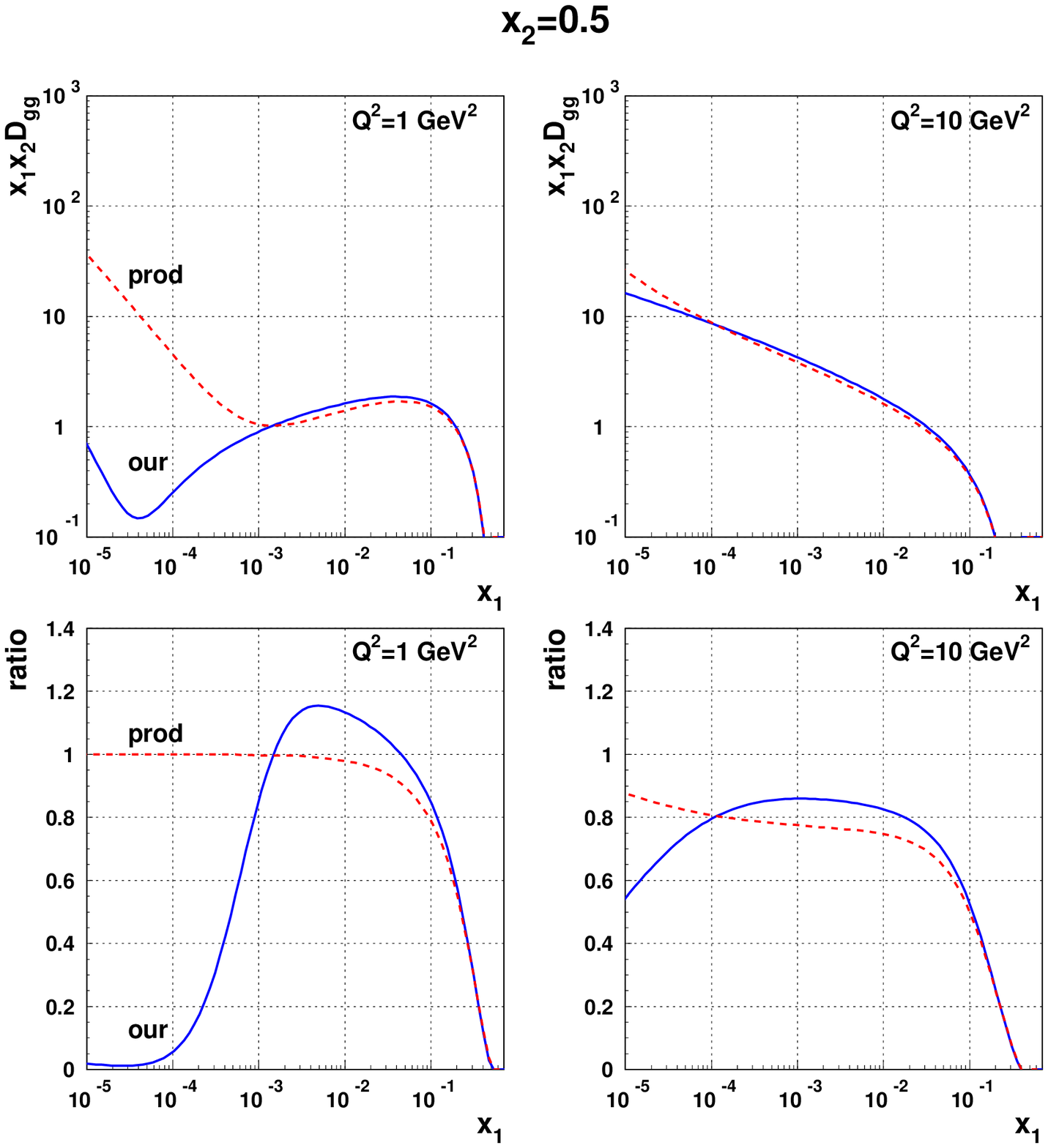}
\caption{The distribution $x_1x_2D_{gg}(x_1,x_2=0.5)$ at  $Q_0^2=1~{\rm GeV}^2$ (left upper panel) and 
$Q^2=10~{\rm GeV}^2$ (right upper panel) and the ratio \eqref{eq:ratio} (lower panels). The solid lines correspond to input (\ref{eq:initialgg}) (our)
while the dashed lines to input (\ref{eq:gaunt}) (prod).
}
\label{fig2}
\end{figure*}


The evolution equations (\ref{eq:twopdfeq})  reduced to the pure gluon case have the following form
\begin{align}\nonumber
\label{eq:twopdfeqgg}
\frac{\partial}{\partial {\ln Q^2}}\, D_{gg}(x_1,x_2,Q)
=\frac{\alpha_s(Q)}{2\pi}
&\Bigg\{\int^{1-x_2}_{x_1} \frac{du}{u} \,{\cal{P}}_{gg}\!\left(\frac{x_1}{u}\right) D_{gg}(u,x_2,Q)
\\\nonumber
&+\int_{x_2}^{1-x_1} \frac{du}{u}\,{\cal{P}}_{gg}\!\left(\frac{x_2}{u}\right)D_{gg}(x_1,u,Q)
\\
&+ \frac{1}{x_1+x_2} \,{P}^R_{gg}\!\left(\frac{x_1}{x_1+x_2}\right) D_{g}(x_1+x_2,Q)\Bigg\}.
\end{align}
where $P^R_{gg}$ is the gluon-to-two gluon splitting function for real emission in the LO approximation. Strictly speaking, such an equation can be 
reasonable approximation for small values of momentum fractions.
Using our numerical program, we solve the above equation with the initial condition (\ref{eq:initialgg}).  We compare our results with 
those obtained from the usually assumed form of the initial conditions \cite{Gaunt:2011xd}, which satisfy  the momentum sum rule only approximately,
\be\label{eq:gaunt}
D_{gg}(x_1,x_2)=D_g(x_1)D_g(x_2)\rho(x_1,x_2)
\ee
where the correlation factor 
\be
\rho(x_1,x_2)=\frac{(1-x_1-x_2)^2}{(1-x_1)^2(1-x_2)^2}\,.
\ee 

The results are shown in Fig.~\ref{fig1} and Fig.~\ref{fig2}. We plot there the double gluon distribution $x_1x_2D_{gg}(x_1,x_2)$ as a function of $x_1$ 
for two values of the scale, initial $Q_0^2=1~{\rm GeV}^2$ and $Q^2=10~{\rm GeV}^2$ (upper panels), for two fixed 
fixed values of $x_2$, small $10^{-2}$ and large $0.5$, respectively, 
 The solid lines show the results obtained from our input
(\ref{eq:initialgg}) while the dashed lines correspond to the input (\ref{eq:gaunt}) with the gluon distribution (\ref{eq:mstwg}). In the lower panels we plot the ratio,
\be
\label{eq:ratio}
{\rm ratio}=\frac{D_{gg}(x_1,x_2)}{D_g(x_1)D_g(x_2)}\,,
\ee
which characterizes factorizability  of the double gluon distribution into a product of two single gluon distributions. 

For both values of $x_2$, the initial double gluon distributions differ significantly for small values of $x_1$, up to $10^{-1}$ for $x_2=10^{-2}$ and
 up to $10^{-3}$ for $x_2=0.5$. However, the QCD evolution equation erases this difference already at the scale $Q^2=10~{\rm GeV}^2$, see the upper
 panels in both figures.
 As we have observed, the initial distribution (\ref{eq:initialgg}) is not factorizable into a product of two single gluon distributions for any values of
 $x_1$ and $x_2$. However, if both momentum fractions are small ($<0.01$,) $D_{gg}$ becomes factorizable  with good accuracy after evolution to the shown value of $Q^2$, see the lower panels in both figures.  A small  breaking of the factorization can be attributed to the non-homogeneous term in the evolution equation 
\eqref{eq:twopdfeqgg}. If one of the two momentum fractions is large, like the shown $x_2=0.5$, this is no longer the case and the factorization is significantly broken
for all values of $x_1$  independent of the values of the evolution scale. We check that for larger values of $Q^2$ that than shown here.
We have to remember, however, that the large $x$ domain has to be supplemented by quarks.

In conclusion, the initial double gluon distribution (\ref{eq:initialgg})
is very different from that proposed so far. However, the QCD evolution equation significantly diminishes this difference at not so high values of the evolution scale $Q^2$. 
\section{Summary}

In this paper we constructed the double gluon distribution $D_{gg}$ from the known single gluon distribution $D_g$, given by the MSTW parameterization, in the pure gluon case. 
The construction is based on the expansion in terms of functions which are 
the Dirichlet distributions. Since
the MSTW distribution has already the form of the sum over Dirichlet distributions of order $2$, 
we postulated the double gluon distribution as a sum over the Dirichlet distributions 
of order $3$   and identified 
the parameters in the two distributions using
the momentum sum rule for this purpose. As a result, we obtained the parameter free double gluon distribution which we evolve with the QCD evolution equation.
We studied the build up of the factorized form of $D_{gg}$ with the increasing evolution scale $Q$. We found that
such a form approximately sets up rather  quickly for small momentum fractions, $x_{1,2}<0.1$. 
As expected, for higher values of $x$,  the factorized form is not valid at all due to the constraint from the momentum sum rule. 

The next step would be to extend this formalism to include the quarks and satisfy the momentum and valence quark sum rules simultaneously. The expansion in terms of the Dirichlet functions can
be constructed also in the case with quarks. The whole formalism is however much more complicated due to the large number of the double parton distribution and therefore the full analysis will be presented in the future publication.
\begin{acknowledgments}
This work was supported by the Polish NCN Grants No.~DEC-2011/01/B/ST2/03915 and DEC-2013/10/E/ST2/00656,  
by the Department of Energy  Grant No. DE-SC-0002145, by the Center 
for Innovation and Transfer of Natural Sciences and Engineering Knowledge in Rzesz\'ow
and by the Angelo Della Riccia foundation. MS 
wishes to thank for hospitality the Penn State University where part of this project was developed.  AMS also thanks the Institute for Nuclear Theory at the University of Washington for its hospitality and the Department of Energy for partial support during the completion of this work.
\end{acknowledgments}


\bibliographystyle{h-physrev4}
\bibliography{mybib}

\begin{thebibliography}{10}

\bibitem{Akesson:1986iv}
Axial Field Spectrometer Collaboration, T.~Akesson {\em et~al.},
\newblock Z.Phys. {\bf C34}, 163 (1987).

\bibitem{Abe:1997bp}
CDF Collaboration, F.~Abe {\em et~al.},
\newblock Phys.Rev.Lett. {\bf 79}, 584 (1997).

\bibitem{Abe:1997xk}
CDF Collaboration, F.~Abe {\em et~al.},
\newblock Phys.Rev. {\bf D56}, 3811 (1997).

\bibitem{Abazov:2009gc}
D0 Collaboration, V.~Abazov {\em et~al.},
\newblock Phys.Rev. {\bf D81}, 052012 (2010), [0912.5104].

\bibitem{Aad:2013bjm}
ATLAS Collaboration, G.~Aad {\em et~al.},
\newblock New J.Phys. {\bf 15}, 033038 (2013), [1301.6872].

\bibitem{Chatrchyan:2013xxa}
CMS Collaboration, S.~Chatrchyan {\em et~al.},
\newblock JHEP {\bf 1403}, 032 (2014), [1312.5729].

\bibitem{Aad:2014rua}
ATLAS Collaboration, G.~Aad {\em et~al.},
\newblock JHEP {\bf 1404}, 172 (2014), [1401.2831].

\bibitem{Shelest:1982dg}
V.~Shelest, A.~Snigirev and G.~Zinovev,
\newblock Phys.Lett. {\bf B113}, 325 (1982).

\bibitem{Zinovev:1982be}
G.~Zinovev, A.~Snigirev and V.~Shelest,
\newblock Theor.Math.Phys. {\bf 51}, 523 (1982).

\bibitem{Ellis:1982cd}
R.~K. Ellis, W.~Furmanski and R.~Petronzio,
\newblock Nucl.Phys. {\bf B212}, 29 (1983).

\bibitem{Bukhvostov:1985rn}
A.~Bukhvostov, G.~Frolov, L.~Lipatov and E.~Kuraev,
\newblock Nucl.Phys. {\bf B258}, 601 (1985).

\bibitem{Snigirev:2003cq}
A.~M. Snigirev,
\newblock Phys. Rev. {\bf D68}, 114012 (2003), [hep-ph/0304172].

\bibitem{Korotkikh:2004bz}
V.~L. Korotkikh and A.~M. Snigirev,
\newblock Phys. Lett. {\bf B594}, 171 (2004), [hep-ph/0404155].

\bibitem{Gaunt:2009re}
J.~R. Gaunt and W.~J. Stirling,
\newblock JHEP {\bf 03}, 005 (2010), [0910.4347].

\bibitem{Blok:2010ge}
B.~Blok, Y.~Dokshitzer, L.~Frankfurt and M.~Strikman,
\newblock Phys.Rev. {\bf D83}, 071501 (2011), [1009.2714].

\bibitem{Ceccopieri:2010kg}
F.~A. Ceccopieri,
\newblock Phys. Lett. {\bf B697}, 482 (2011), [1011.6586].

\bibitem{Diehl:2011tt}
M.~Diehl and A.~Schafer,
\newblock Phys. Lett. {\bf B698}, 389 (2011), [1102.3081].

\bibitem{Gaunt:2011xd}
J.~R. Gaunt and W.~J. Stirling,
\newblock JHEP {\bf 1106}, 048 (2011), [1103.1888].

\bibitem{Ryskin:2011kk}
M.~Ryskin and A.~Snigirev,
\newblock Phys.Rev. {\bf D83}, 114047 (2011), [1103.3495].

\bibitem{Bartels:2011qi}
J.~Bartels and M.~G. Ryskin,
\newblock 1105.1638.

\bibitem{Blok:2011bu}
B.~Blok, Y.~Dokshitzer, L.~Frankfurt and M.~Strikman,
\newblock Eur.Phys.J. {\bf C72}, 1963 (2012), [1106.5533].

\bibitem{Diehl:2011yj}
M.~Diehl, D.~Ostermeier and A.~Schafer,
\newblock JHEP {\bf 1203}, 089 (2012), [1111.0910].

\bibitem{Luszczak:2011zp}
M.~Luszczak, R.~Maciula and A.~Szczurek,
\newblock Phys. Rev. {\bf D85}, 094034 (2012), [1111.3255].

\bibitem{Manohar:2012jr}
A.~V. Manohar and W.~J. Waalewijn,
\newblock Phys.Rev. {\bf D85}, 114009 (2012), [1202.3794].

\bibitem{Ryskin:2012qx}
M.~Ryskin and A.~Snigirev,
\newblock Phys.Rev. {\bf D86}, 014018 (2012), [1203.2330].

\bibitem{Gaunt:2012dd}
J.~R. Gaunt,
\newblock JHEP {\bf 1301}, 042 (2013), [1207.0480].

\bibitem{Blok:2013bpa}
B.~Blok, Y.~Dokshitzer, L.~Frankfurt and M.~Strikman,
\newblock Eur.Phys.J. {\bf C74}, 2926 (2014), [1306.3763].

\bibitem{vanHameren:2014ava}
A.~van Hameren, R.~Maciula and A.~Szczurek,
\newblock Phys. Rev. {\bf D89}, 094019 (2014), [1402.6972].

\bibitem{Maciula:2014pla}
R.~Maciula and A.~Szczurek,
\newblock Phys. Rev. {\bf D90}, 014022 (2014), [1403.2595].

\bibitem{Snigirev:2014eua}
A.~Snigirev, N.~Snigireva and G.~Zinovjev,
\newblock Phys.Rev. {\bf D90}, 014015 (2014), [1403.6947].

\bibitem{Golec-Biernat:2014nsa}
K.~Golec-Biernat and E.~Lewandowska,
\newblock Phys.Rev. {\bf D90}, 094032 (2014), [1407.4038].

\bibitem{Gaunt:2014rua}
J.~R. Gaunt, R.~Maciula and A.~Szczurek,
\newblock Phys. Rev. {\bf D90}, 054017 (2014), [1407.5821].

\bibitem{Harland-Lang:2014efa}
L.~A. Harland-Lang, V.~A. Khoze and M.~G. Ryskin,
\newblock J. Phys. {\bf G42}, 055001 (2015), [1409.4785].

\bibitem{Blok:2014rza}
B.~Blok and M.~Strikman,
\newblock Eur. Phys. J. {\bf C74}, 3214 (2014), [1410.5064].

\bibitem{Maciula:2015vza}
R.~Maciula and A.~Szczurek,
\newblock 1503.08022.

\bibitem{Kirschner:1979im}
R.~Kirschner,
\newblock Phys.Lett. {\bf B84}, 266 (1979).

\bibitem{Golec-Biernat:2014bva}
K.~Golec-Biernat and E.~Lewandowska,
\newblock Phys.Rev. {\bf D90}, 014032 (2014), [1402.4079].

\bibitem{Broniowski:2013xba}
W.~Broniowski and E.~Ruiz~Arriola,
\newblock Few Body Syst. {\bf 55}, 381 (2014), [1310.8419].

\bibitem{Martin:2009iq}
A.~Martin, W.~Stirling, R.~Thorne and G.~Watt,
\newblock Eur.Phys.J. {\bf C63}, 189 (2009), [0901.0002].

\bibitem{Manohar:2012pe}
A.~V. Manohar and W.~J. Waalewijn,
\newblock Phys.Lett. {\bf B713}, 196 (2012), [1202.5034].

\bibitem{Blok:2012jr}
B.~Blok, M.~Strikman and U.~A. Wiedemann,
\newblock Eur. Phys. J. {\bf C73}, 2433 (2013), [1210.1477].

\end{thebibliography}

\end{document}